\documentclass{emulateapj}
\usepackage{apjfonts}
\usepackage{epsf}
\begin{document}
\slugcomment{}
\shortauthors{J. M. Miller et al.}
\shorttitle{GRS 1915$+$105}

\title{The Accretion Disk Wind in the Black Hole GRS 1915+105}

\author{J.~M.~Miller\altaffilmark{1},
J. Raymond\altaffilmark{2}, 
A. C. Fabian\altaffilmark{3}, 
E. Gallo\altaffilmark{1}
J. Kaastra\altaffilmark{4,5}, 
T. Kallman\altaffilmark{6},
A. L. King\altaffilmark{7,8,9},
D. Proga\altaffilmark{10}, 
C. S. Reynolds\altaffilmark{11}
A. Zoghbi\altaffilmark{1}
}
 
\altaffiltext{1}{Department of Astronomy, University of Michigan, 1085
  South University Avenue, Ann Arbor, MI 48109-1107, USA,
  jonmm@umich.edu}

\altaffiltext{2}{Harvard-Smithsonian Center for Astrophysics, 60 Garden Street, Cambridge, MA 02138, USA}

\altaffiltext{3}{Institute of Astronomy, University of Cambridge,
  Madingley Road, Cambridge CB3 OHA, UK}

\altaffiltext{4}{SRON Netherlands Institute for Space Research, Sorbonnelaan 2, 3584 CA Utrecht, NL}

\altaffiltext{5}{Department of Physics and Astronomy, Universiteit
  Utrecht, PO Box 80000, 3508 TA Utrecht, NL}

\altaffiltext{6}{NASA Goddard Space Flight Center, Code 662, Greedbelt, MD 20771, USA}

\altaffiltext{7}{Department of Physics, Stanford University, 382 Via
  Pueblo Mall, Stanford, CA, 94305}

\altaffiltext{8}{Einstein Fellow}

\altaffiltext{9}{Kavli Fellow}

\altaffiltext{10}{Department of Physics, University of Nevada, Las
  Vegas, Las Vegas, NV 89154, USA}

\altaffiltext{11}{Department of Astronomy, University of Maryland, College Park, MD 20742-2421, USA}

\keywords{accretion disks -- black hole physics -- X-rays: binaries}

\label{firstpage}

\begin{abstract}
We report on a 120~ks {\it Chandra}/HETG spectrum of the black hole
GRS 1915$+$105.  The observation was made during an extended and
bright soft state in June, 2015.  An extremely rich disk wind
absorption spectrum is detected, similar to that observed at lower
sensitivity in 2007.  The very high resolution of the {\it
  third-order} spectrum reveals four components to the disk wind in
the Fe K band alone; the fastest has a blue-shift of $v = 0.03c$.
Broadened re-emission from the wind is also detected in the
first-order spectrum, giving rise to clear accretion disk P Cygni
profiles.  Dynamical modeling of the re-emission spectrum gives wind
launching radii of $r \simeq 10^{2-4}~{\rm GM/c^{2}}$.  Wind density
values of $n \simeq 10^{13-16}~ {\rm cm}^{-3}$ are then required by
the ionization parameter formalism.  The small launching radii, high
density values, and inferred high mass outflow rates signal a role for
magnetic driving.  With simple, reasonable assumptions, the wind
properties constrain the magnitude of the emergent magnetic field to
be $B\simeq 10^{3-4}$~Gauss if the wind is driven via
magnetohydrodynamic (MHD) pressure from within the disk, and $B\simeq
10^{4-5}$~Gauss if the wind is driven by magnetocentrifugal
acceleration.  The MHD estimates are below upper
limits predicted by the canonical $\alpha$-disk model (Shakura \&
Sunyaev 1973).  We discuss these results in terms of fundamental disk
physics and black hole accretion modes.
\end{abstract}

\section{Introduction}
Disk accretion onto stellar-mass black holes at high Eddington
fractions is dominated by thermal emission from an accretion disk.
This blackbody-like emission can be characterized extremely well with
just two parameters: flux, and color temperature.  This simplicity
enables efficient traces of disk temperature across orders of
magnitude in flux (e.g. Reynolds \& Miller 2013), but it hides the
underlying physics that mediate the accretion process.

The magneto-rotational instability (MRI; Balbus \& Hawley 1991)
appears to supply the degree of effective viscosity required to drive
disk accretion.  Alternatively, gas might escape along poloidal
magnetic field lines, transferring angular momentum and allowing mass
transfer through the disk (e.g., Blandford \& Payne 1982).  Accretion
in FU Ori and T Tauri systems suggests that these mechanisms might
operate concurrently: thermal continuum emission signals a viscous
disk, but the profiles observed in line spectra signal a rotating disk
wind (e.g., Calvet, Kenyon, \& Hartmann 1993).  If young stellar disks
are any guide, line spectra and winds may hold the key to
understanding disks around black holes.

Miller et al.\ (2015) recently examined the richest disk winds found
in {\it Chandra}/HETG spectra of the stellar-mass black holes
4U~1630$-$472, H~1743$-$322, GRO J1655$-$40, and GRS 1915$+$105.  The
observations were obtained in the disk--dominated ``high/soft'' or
``thermal dominant'' state (see, e.g., Remillard \& McClintock 2006).
The Fe K band was found to require 2--3 velocity/ionization
components when fit with new, self-consistent XSTAR photoionization
models.  The spectra also require re-emission from the dense absorbing
gas, and the emission is broadened by a degree that is loosely
consistent with Keplerian orbital motion at the photoionization
radius.  This provides a means of estimating wind launching radii,
densities, outflow rates, kinetic power, and driving mechanisms.

The observation of GRS 1915$+$105 treated in Miller et al.\ (2015) was
obtained during an extended soft state in 2007 (also see Ueda et
al.\ 2009, Neilsen \& Lee 2009).  In the spring of 2015, GRS
1915$+$105 was observed to be in a long decline in the SWIFT/BAT
(15--50~keV).  The source flux eventually became consistent with zero
in single visits, indicating that GRS~1915$+$105 had again locked into
a steady soft state.  We therefore triggered an approved {\it Chandra}
target of opportunity (TOO) observation of GRS 1915$+$105.\\

\section{Observations and Reduction}
GRS 1915$+$105 was observed for 120~ks starting on 2015 June 9 at
15:30:59 UT ("obsid" 16711), using the HETGS.  Owing to the high flux
level expected in the {\it Chandra} band, the ACIS-S array was
read-out in "continuous clocking" mode to prevent photon pile-up,
reducing the nominal frame time to just 2.85~ms.  A 100-column "gray"
filter was applied across the full height of the S3 chip, windowing
the zeroth-order aimpoint.  Within this window, one in 10 events was
telemetered; this enables the wavelength grid to be reconstructed
while preserving telemetry.

CIAO version 4.7 and the associated current calibration files were
used to generate spectral files and responses.  The tool
"add\_grating\_orders" was used to combine opposing spectra from the
first and third orders.  Redistribution matrices were created
using the tool "mkgrmf" and the ancillary responses were created
using "mkgarf".

The Fe K band traces the mostly highly ionized gas, likely originating
closest to the black hole.  Following Miller et al.\ (2015), we
proceed to only analyze the combined third-order HEG spectrum in the
5--8 keV band (above 8~keV, order sorting becomes difficult and stray
flux enters the extraction region), and the combined first-order HEG
spectrum over the 5--10~keV band.

\section{Analysis and Results}
Models were fit to the data using XSPEC version 12.8.2 (Arnaud 1996).
In all of the fits, "Churazov" weighting (Churazov et al.\ 1996) was
adopted.  All of the errors reported in this work are $1\sigma$
confidence intervals.

An equivalent neutral hydrogen column density of $N_{\rm ISM} =
4.0\times 10^{22}~ {\rm cm}^{-2}$ was assumed in all fits and modeled
using ``tbabs'' (Wilms et al.\ 2000).  Fits to the first-order
spectrum confirmed that the continuum can be well described with disk
blackbody (Mitsuda et al.\ 1984) and power-law components.  The $kT
\simeq 1.6$~keV thermal emission dominates over the steep ($\Gamma
\simeq 2.8$) power-law (see Table 1).  This characterization of the
first-order spectrum gives an unabsorbed flux of $F \simeq 4.1\times
10^{-8}~ {\rm erg}~ {\rm cm}^{-2}~ {\rm s}^{-1}$ in the 0.5--30.0~keV
band, corresponding to a luminosity of $L \simeq 3.7\times 10^{38}~
{\rm erg}~ {\rm s}^{-1}$ for $d = 8.6$~kpc (Reid et
al.\ 2014).

In order to characterize the the wind, we generated a grid of
photoionized line spectra using XSTAR version 2.2.1bo8 (see, e.g.,
Kallman et al.\ 2009).  This (newest) version of the code includes the
effects of resonant scattering when calculating emission spectra, in
addition to recombination, fluorescence, and collisional ionization.
A blackbody input spectrum with $kT = 1.6$~keV and $L = 3.7\times
10^{38}~ {\rm erg}~ {\rm s}^{-1}$ was assumed.  Following Miller et
al.\ (2015), a turbulent velocity of $300~ {\rm km}~ {\rm s}^{-1}$,
solar abundances (except for iron, fixed at twice the solar value as
per Lee et al.\ 2002), a covering factor of $\Omega/4\pi = 0.5$, and a
gas density of $n = 10^{14}~ {\rm cm}^{-3}$ were assumed.  Via the
"xstar2xspec" function, 400 photoionization models were created,
spanning a range of $6.0\times10^{21}~ {\rm cm}^{-2} \leq N \leq
6.0\times10^{23}$, and $3.0 \leq {\rm log}(\xi) \leq 6.0$.  The
resolution of the models was set to give 100,000 spectral bins.

Photoionized absorption was included in the overall model as a
multiplicative component acting upon the continuum.  It was
characterized in terms of a column density, ionization parameter, and
velocity shift (${\rm N}_{H, wind}$, $\xi = L/nr^{2}$, and $v/c$).
Self-consistency demands an additive re-emission component for each
absorption component.  The column density and ionization parameter
were linked between the absorption and re-emission components in each
zone.  The emission component carries a flux normalization, given by
$K = (\Omega/4\pi)~L_{38} / d_{kpc}^{2}$ (where $L_{38}$ is the
luminosity in units of $10^{38}~{\rm erg}~{\rm s}^{-1}$ and $d_{kpc}$
is the distance in units of kpc).  Normalization values were
restricted to the 0.1--10.0 range.  The re-emission spectra are
consistent with a gas at zero velocity shift, per emission from a
broad range of azimuth in a cylindrical geometry.  Emission velocity
shifts were then fixed at zero.

The re-emission is broadened, and is expected to have a non-Gaussian
shape if it originates from several hundred or several thousand radii.
We therefore convolved each re-emission component with the ``rdblur''
function (Fabian et al.\ 1989).  This Schwarzschild function differs
negligibly from the anticipated (see, e.g., McClintock et al.\ 2006,
Miller et al.\ 2013) near-maximal Kerr potential far from the black
hole, and its range extends to the launching radii whereas newer Kerr
blurring functions do not.  The inclination was jointly determined
between all zones (restricted to the $60^{\circ} \leq \theta \leq
80^{\circ}$ range).  Tests revealed that a constant density emissivity
profile ($r^{-2}$) gave the best fits, with the outer radius fixed at
a multiple of the inner radius for all zones.  A value of 5.0 gave the
lowest fit statistic ($R_{out} = 5.0\times R_{in}$), and values
reported in Table 1 reflect fits made with this scheme.  Last, a
Gaussian emission feature with an energy constrained to lie in the
6.40--6.43~keV range (Fe I-XVII) was included to account for any
distant, low-ionization emission.

Initial fits were made to the first and third-order spectra
separately.  A model with three zones is able to reproduce the lines
in the third-order spectrum below 7~keV, and even an H-like absorption
doublet blue-shifted by approximately 0.01~c, up to 7.05~keV.
However, a weak H-like doublet is apparent at approximately 7.2~keV.
We therefore included a fourth zone in fits to the third-order
spectrum, measuring a blue-shift of $v = 0.0305(4)~c$.

We next made joint fits to the first-order and third-order spectra.
The continuum in the first-order spectrum is robust and reported in
Table 1, but the continuum in the third-order is affected by stray
continuum flux, so the continuum parameters and column densities are
not linked between the two spectra.  The first-order spectrum is
generally more sensitive and the parameters of the third-order wind model
were tied to those for the first-order spectrum, with the exception of
the third and fourth zones, which carry higher velocity shifts.  For
those zones, the velocity is determined by the third-order spectrum.

The best-fit model is detailed in Table 1.  The fits are shown in
Figures 1 and 2.  Figure 3 illustrates that the re-emission spectrum
is broadened.  Last, Figure 4 compares the 2007 and 2015
spectra of the high/soft state in GRS 1915$+$105.  Overall, a good fit
is achieved, with $\chi^{2}/\nu = 1951.2/1572 = 1.241$.  Each zone is
required by the data, as measured by the F-statistic.  The best
two-zone model is an enormous ($>>8\sigma$) improvement over the best
single-zone model.  The best three-zone model is a $5\sigma$
improvement over any fit with just two zones.  The addition of the
fourth zone is a $3\sigma$ improvement over any three-zone model.
Re-emission from Zone 3 is not highly broadened, suggesting a more
distant origin.  The re-emission from Zone 4 is not
well-constrained, and was fixed at a representative value of $K =
0.5$.  In general, the re-emission normalizations should be regarded
as relative scaling factors, since they are partially affected by
small continuum disparities between the first and third-order spectra.

The slower components of the wind (Zones 1 and 2 in Table 1) have
lower ionizations and higher densities; in contrast, the faster
components (Zones 3 and 4 in Table 1) have higher ionizations, and
lower densities.  This may be consistent with acceleration within the
absorbing region.  However, the fastest component - Zone 4 - appears
to originate at {\it small} radii.  The wind is likely inconsistent
with a homogeneous flow with uniform acceleration, but it may be
consistent with a complex flow with multiple stream lines.

Table 1 also gives estimates of the gas density, mass outflow rate,
kinetic power, and filling factor in each wind zone.  The gas density
was obtained through the ionization parameter ($\xi = L/nr^{2}$),
utilizing the radius implied by the broadening of the re-emission.
The mass outflow rate is estimated without assuming a density by
utilizing $L/\xi = nr^{2}$, giving $\dot{M}_{wind} = \Omega \mu m_{p}
v (L/\xi)$.  The kinetic power is then just $L_{kin} = 0.5
\dot{M}_{wind} v^{2}$.  It is notable that the outflow rate is
comparable to the accretion rate in the inner disk in Zones 2, 3, and 4.
Summing all zones, the kinetic power of the wind is nominally about
0.1\% of the radiative power.  The thickness of each wind zone can be
estimated via $\Delta r = N/n$, and the volume filling factor is then
given by $f = \Delta r/r$.  Table 1 gives estimates of $f$ for each
zone.  Depending on the geometry, it may or may not be appropriate to
multiply the mass outflow rates and kinetic power by $f$, reducing
both quantities.  The best fits are obtained when $\Delta r/r = 5$ in
the blurring function; this may indicate that the filling factor is
not very small.

Radiative line driving is only effective for ${\rm log}(\xi) \leq 3$
(Proga 2003); the wind cannot be driven in this manner.  Thermal
driving is effective outside of the Comptonization radius, $R_{C} =
10^{10}\times (M_{BH}/(M_{\odot}) / T_{C,8}~{\rm cm}$ (where $T_{C,8}$
is the Compton temperature in units of $10^{8}$~K; Begelman et
al.\ 1983).  Other work suggests that winds can be driven from
0.1~$R_{C}$ (Woods et al.\ 1996; also see Proga \& Kallman 2002).
Approximating the Compton temperature with the disk temperature
(reasonable under equilibrium conditions), $R_{C} \simeq 5\times
10^{11}$~cm, or $3.3\times 10^{5}~{\rm GM/c^{2}}$.  Thus, $R_{C}$ is
2--3 orders of magnitude larger than implied by the broadening of the
re-emission spectra of Zones 1, 2, and 4 (see Table 1).  The wind
densities and outflow rates also exceed estimates from thermal driving
models, though such models are evolving (e.g., Higginbottom \& Proga
2015).  Emission line broadening might arise via turbulent motions in
thermal winds (e.g., Sim et al.\ 2010).  However, this would lead to
similar line widths in emission and absorption, whereas the emission
lines are much broader than the absorption lines in our best-fit model
(see Figure 3).

If the wind is driven by MHD pressure (e.g., Proga et al.\ 2003), the
magnetic pressure must equal or exceed the gas pressure in the wind,
meaning that $B^{2}/8\pi = 2nkT$, or $B = \sqrt{16\pi nkT}$.  If
magnetocentrifugal driving (Blandford \& Payne 1982) dominates, the
magnetic field pressure must equal or exceed the ram pressure of the
wind (else the poloidal field geometry will break down), giving
$B^{2}/8\pi = 0.5 \rho v_{tot}^{2}$, or $B = \sqrt{4\pi \mu m_{p} n
  v_{tot}^{2}}$ (where $v_{tot}$ includes radial and azimuthal
velocities).  We ran new XSTAR models assuming the derived densities
to ascertain the temperature within the wind; these temperatures and
the magnetic field magnitudes are listed in Table 1.  Using equation
2.19 in Shakura \& Sunyaev (1973), limits on the magnetic field
strength expected in the $\alpha$-disk prescription were calculated;
these are also listed in Table 1.

\section{Discussion and Conclusions}
We have obtained a sensitive {\it Chandra}/HETGS observation of the
black hole GRS 1915$+$105, in an extended soft state.  The spectrum
shows strong absorption lines in the Fe K band, and weaker, broadened
emission lines, together creating accretion disk P Cygni profiles (see
Figures 1--4).  Four distinct wind zones are detected in fits to the
first and third-order spectra; the fastest has a blue shift of
$v = 0.03c$.  The broadening of the re-emission components of the wind
permits measurements of its launching radius ($R \simeq 10^{2-4}~{\rm
  GM/c^{2}}$), which in turn allow for estimates of the wind density
($n \simeq 10^{13-16}~ {\rm cm}^{-3}$) and other parameters.  The wind is likely
driven by magnetic processes, connecting the wind to fundamental
aspects of disk accretion, including momentum and mass transfer.  

Though luminosities can be difficult to estimate even for X-ray
binaries, we observed GRS 1915$+$105 at an apparent luminosity of
$L = 0.28~ L_{Edd}$, assuming the most recent estimates of its distance
and mass.  This is significant because simulations suggest that
standard {\it thin} accretion disks operate at this Eddington fraction
(e.g., Shafee et al.\ 2008, Reynolds \& Fabian 2008).  Simulations
find that winds might be Compton-thick very close to the disk, but do
not remain so as the gas moves upward and outward (Chakravorty et
al.\ 2016).  It is possible but unlikely that a Compton-thick,
super-Eddington flow was observed in GRS 1915$+$105; all of the column
densities measured in Table 1 are significantly below ${\rm N}_{\rm H}
= 10^{24}~{\rm cm}^{-2}$, and the total kinetic power of the wind is
only a small fraction ($\simeq 0.1\%$) of the radiated luminosity.

Wind rotation might nominally favor magnetocentrifugal driving
(Blandford \& Payne 1982), but rotation is not necessarily unique to
this mechanism.  Moreover, a low filling factor may be required in
order to hold the mass outflow rate below the accretion inflow rate
(else the wind transfers more angular momentum than the disk can
supply; see, e.g., Reynolds 2012).  Velocity profiles can shed
additional light: if an MHD wind is a momentum-conserving flow, then
azimuthal velocity should decrease linearly ($v \propto R^{-1}$);
however, in the magnetocentrifugal case, azimuthal velocity increases
linearly with radius ($v \propto R$).  Associating observed
blue-shifts with local escape speeds gives a poor radius estimate
(effectively an upper limit).  Our results reveal no trend between
rotational broadening (traced by the radius estimate from blurring
with ``rdblur'') and radius (poorly traced by outflow speed).  This is
consistent with a combination of MHD and magnetocentrifugal driving;
the wind would then fail to conserve momentum and transfer some away
from the disk, aiding mass transfer through the disk.

The magnetic field estimates made assuming that MHD pressure drives
the wind are safely below the limits predicted by $\alpha$-disk models
(Shakura \& Sunyaev 1973; see Table 1).  This may offer new support
for the basic framework of $\alpha$-disk models, at least in the inner
disk in stellar-mass black holes.  Moreover, the magnetic energy flux
predicted in MRI disk simulations by Miller \& Stone (2000) appears
sufficient to power the wind launched in the soft state of GRS
1915$+$105.  The equatorial nature of this disk wind (and others) is
also consistent with simulations of MHD winds; indeed, the inferred
wind parameters (such as density) closely resemble some values
obtained in Proga et al.\ (2003).  Chakravorty et al.\ (2016)
simulated MHD winds in X-ray binaries and found that magnetic winds
are possible at small radii for high densities and high ionizations;
this appears to confirm AGN-focused studies of MHD flows (e.g.,
Fukumura et al.\ 2014, 2015).  In the broadest sense, magnetic field
constraints from disk winds may mark a turning point in our ability to
probe fundamental disk physics, and the start of closer comparisons
between observations and simulations.

Observations indicate that winds and jets are anti-correlated by
spectral state (e.g., Miller et al.\ 2006a,b,2008; Neilsen \& Lee
2009, King et al.\ 2012, Ponti et al.\ 2012).  This may also indicate
a link between disk properties, magnetic field configurations, and
outflow modes.  Begleman et al. (2015) have proposed an association
between the spectral state and the structure of the dynamo based on an
analytic description of magnetically dominated disks supported by
shearing box simulations (Salvesen et al. 2016).  The close similarity
in wind properties in soft states of GRS 1915$+$105 {\it separated by
  eight years} signals an even closer relationship between the state
of the disk and the nature of the wind that is launched (see Figure
4).  An intriguing possibility is that winds may act to partly
regulate the operation of the disk.  Obtaining sensitive spectra at a
high cadence could detect changes in the wind and disk continuum, and
determine which geometry leads variations.

We thank the anonymous referee for suggestions that improved this paper.

\clearpage

\begin{figure}
  \hspace{0.25in}
  \includegraphics[scale=0.6,angle=-90]{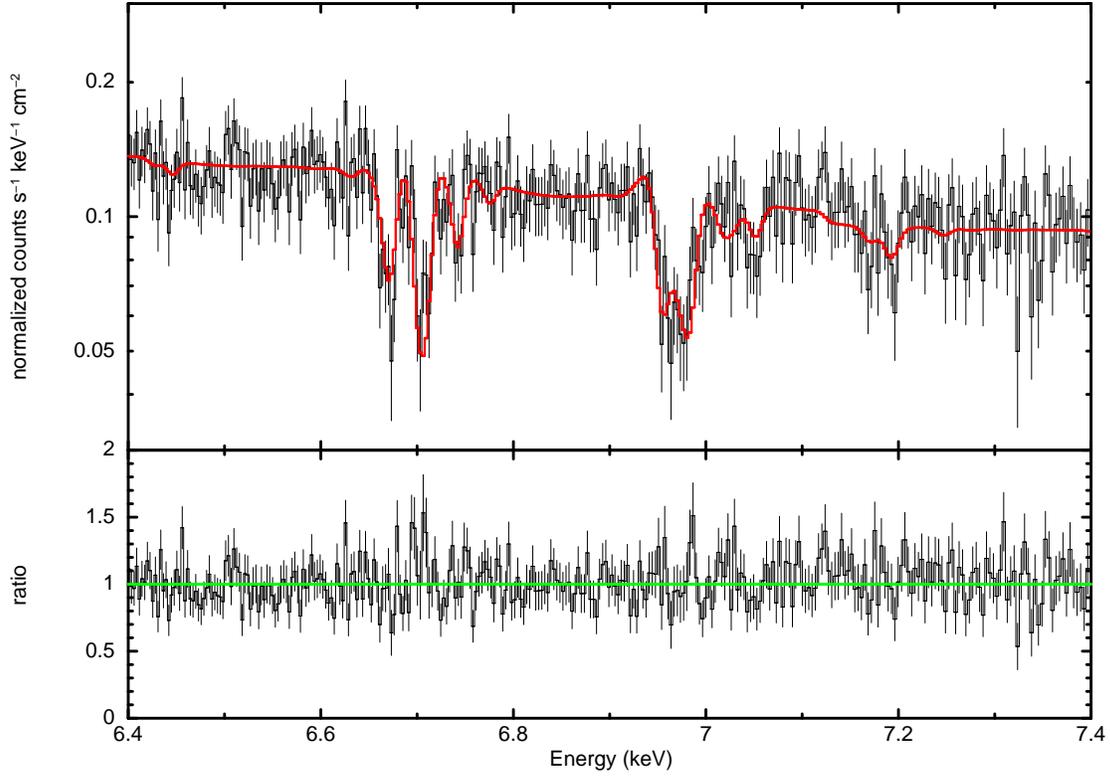}
\figcaption[t]{\footnotesize The third-order spectrum of GRS
  1915$+$105.  The best-fit model
  based on joint fits to the first and third-order spectra is shown in
  red (see Table 1).  Four photoionization zones with paired
  absorption and re-emission are required to fit the data.  The
  He-like Fe XXV line is resolved into intercombination and resonance
  lines (rest frame energy: 6.700~keV).  Instances of H-like Fe XXVI
  absorption lines close to the rest-frame value of 6.970~keV, and
  blue-shifted up to 7.05~keV and 7.2~keV are apparent.  The Fe XXVI
  line shape is a doublet owing to the expected spin-orbit splitting
  in the H-like atom.}
\vspace{0.25in}
\end{figure}
\medskip

\begin{figure}
  \hspace{0.25in}
  \includegraphics[scale=0.6,angle=-90]{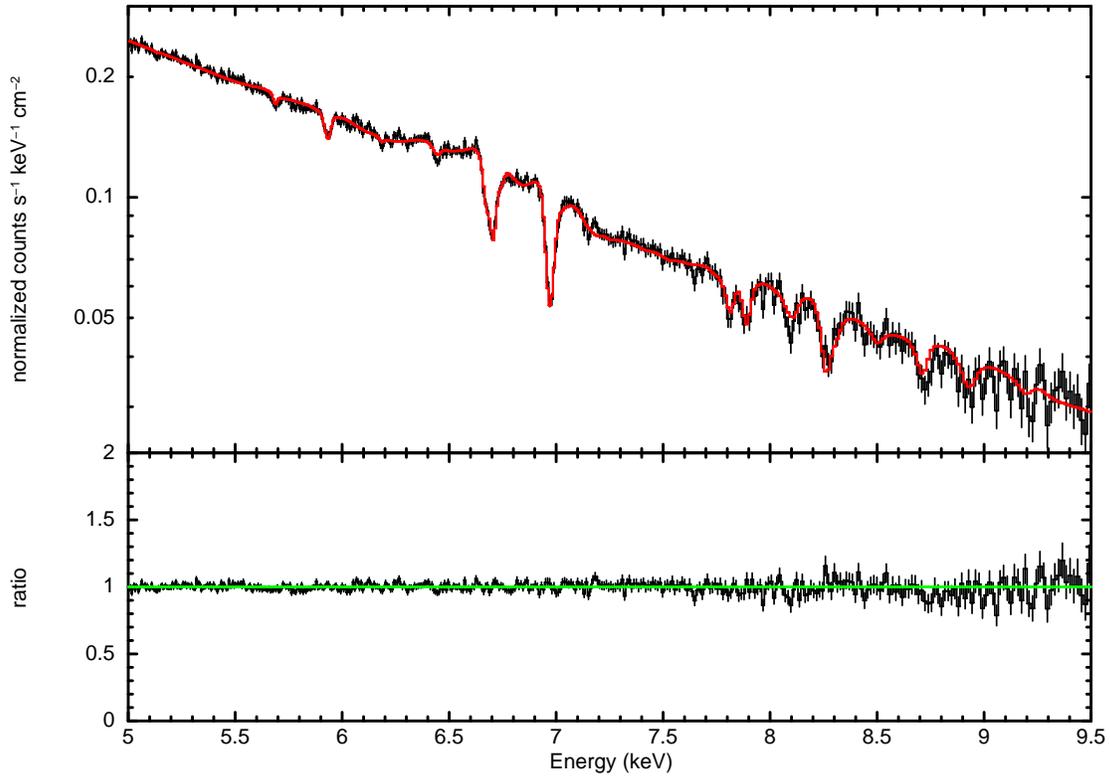}
\figcaption[t]{\footnotesize The first-order spectrum of GRS
  1915$+$105.  The best-fit
  model is shown in red (see the text, and Table 1).}
\end{figure}
\medskip

\clearpage

\begin{figure}
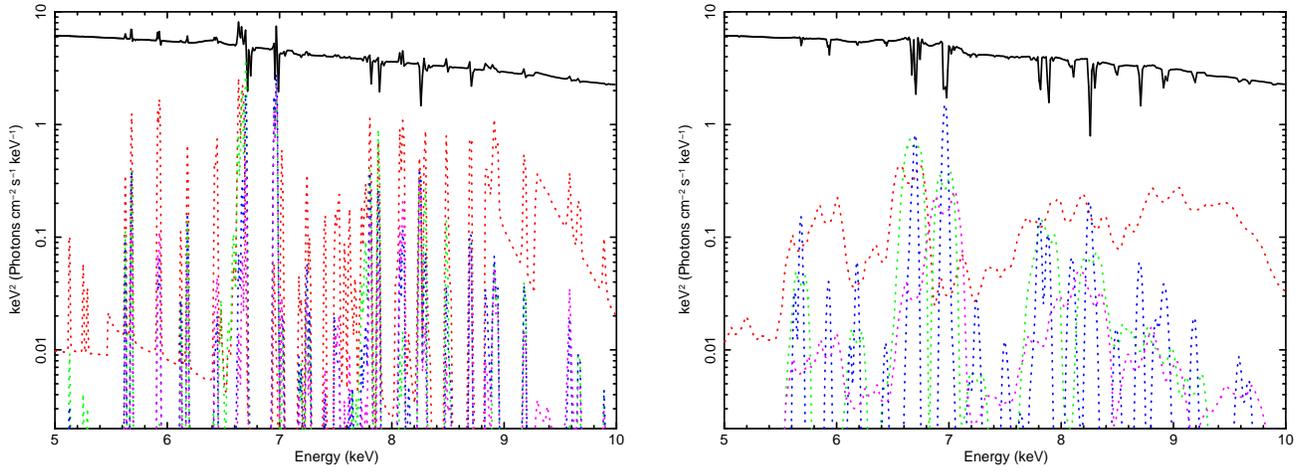

\includegraphics[scale=0.35,angle=-90]{f3a.ps}
\includegraphics[scale=0.35,angle=-90]{f3b.ps}
  \figcaption[t]{\footnotesize Dynamical broadening plays an important
    role in shaping the re-emission spectrum of the wind.  Zones 1--4
    are depicted in red, green, blue, and magenta, respectively.  {\bf
      Left}: Dynamical broadening has been removed from the model in
    Table 1, yielding wind re-emission lines that are much sharper
    than the data.  {\bf Right}: The spectral model detailed in Table
    1, with the blurring required by the data.}
\end{figure}
\medskip

\clearpage

\begin{figure}
\hspace{0.5in}
  \includegraphics[scale=0.8]{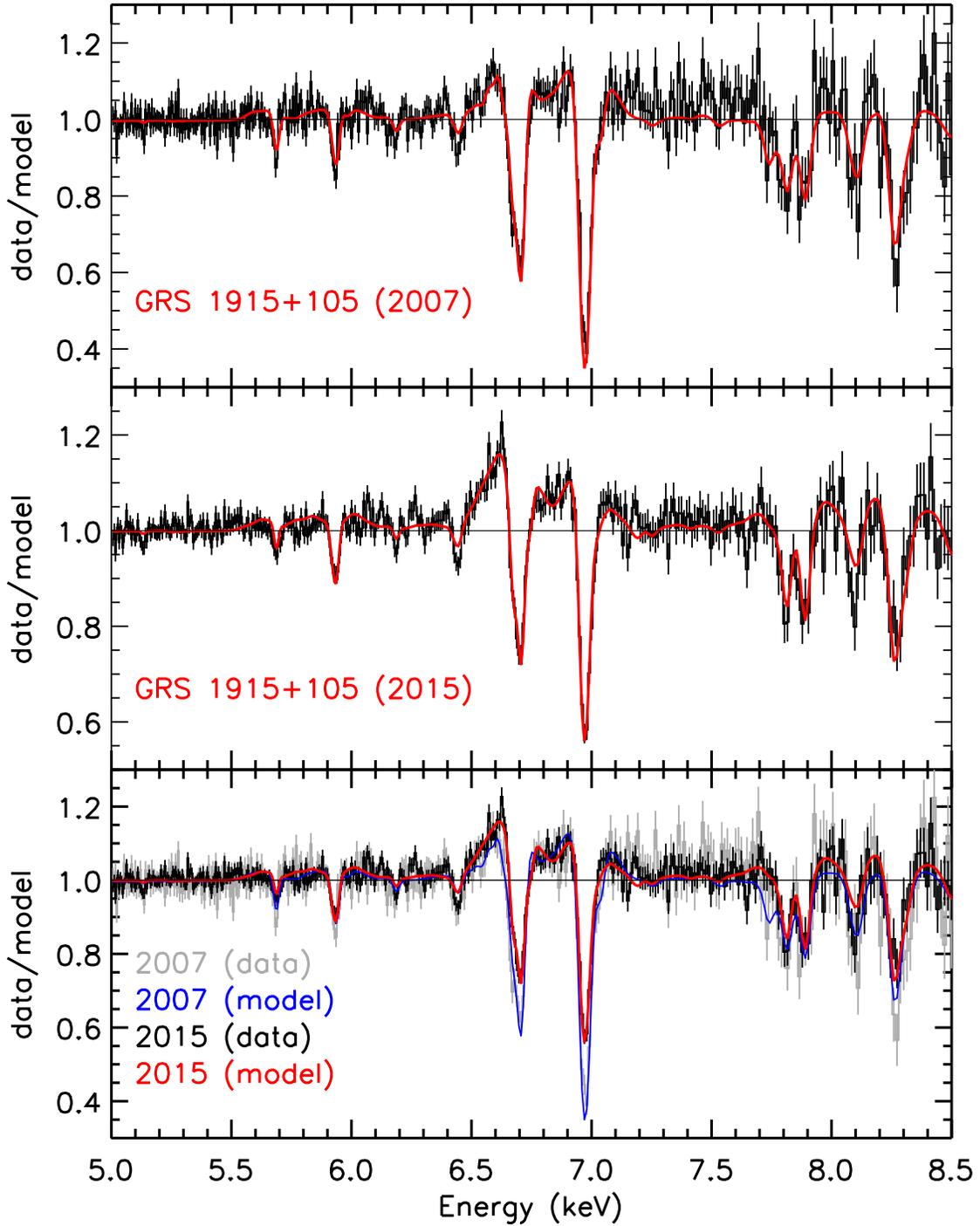}
  \figcaption[t]{\footnotesize First-order HEG spectra of GRS
    1915$+$105 from the soft states observed in 2007 and 2015.  The
    spectra are shown as a ratio to the best-fit continuum, with the
    best-fit wind spectrum plotted through the ratio.  In this
    representation, the presence of accretion disk P Cygni line
    profiles is clear.  {\bf Top}: Data from the 2007 soft
    state, with the model of Miller et al.\ (2015).  {\bf Middle}:
    Data from the 2015 soft state, with the corresponding best-fit
    model (see Figure 2 and Table 1).  {\bf Bottom}: The 2007 and 2015
    data and respective models are plotted together.  The similarity
    of the spectra suggests that a particular wind geometry
    manifests in soft states.}
\end{figure}
\medskip

\clearpage

\begin{table}[t]
\caption{Spectral Fitting Results and Derived Wind Parameters}
\begin{footnotesize}
\begin{center}
\begin{tabular}{llllll}
\tableline
\tableline
Parameter & Zone 1 & Zone 2 & Zone 3 & Zone 4 &  Continuum  \\
\tableline
${\rm N}_{\rm H}~(10^{22}~{\rm cm}^{-2})$ & $30(2)$ & $0.65(5)$ & $1.0(3)$ & $2.5^{+2.5}_{-1.2}$ & --  \\
log($\xi$)     &  $4.04(2)$    & $3.87(5)$    & $4.5(1)$  &  $4.7(5)$  & --       \\
$v/c~(10^{-3})$     &  $-0.70(5)$  & $-6.2(3)$   & $-11.0(5)$  & $-30.5(5)$    &   --        \\
$R~({\rm GM/c^{2}})$      &  $850^{+250}_{-50}$ & $3000^{+600}_{-400}$  & $30,000_{-2000}$ & $1200^{+300}_{-300}$   &  --     \\
$\theta$~(deg)      &  $60^{+7}$ & $60^{+7}$ & $60^{+7}$   &  $60^{+7}$    & --  \\
emis.~norm.     &  $0.10^{+0.01}$ &  $0.21(2)$ &  $0.38(7)$ &   $0.5*$       & --  \\
\tableline
$n~ (10^{15}~ {\rm cm}^{-3})$ & 20(5)  & 2.4(6)  & 0.006(1) &  2.2(3)  & --  \\
$\dot{\rm M}_{wind}~ (10^{18}~{\rm g}~{\rm s}^{-1})$  & 0.73(6) & 10(1) & 4.0(9) & 8(8) & --  \\
$\dot{\rm M}_{wind}~ (10^{-8}~{\rm M}_{\odot}~ {\rm yr}^{-1})$  & 1.2(1) & 15(2) & 6(2) & 1(1) & --  \\ 
$\dot{\rm M}_{wind}/\dot{\rm M}_{accr.}$ &  0.18(2) & 2.3(3) & 1.0(3) & 2(2) & -- \\
$L_{kin} (10^{34}~ {\rm erg}~ {\rm s}^{-1})$ & 0.016(2) & 17(2) & 22(5) & 30(30) & -- \\
$f~(10^{-2})$                             & 1.3(4)  & 0.06(2) & 4(1)  & 0.1(1) & -- \\                                    
${\rm log}({\rm T}_{wind})$ & 6.5(2) & 6.0(2) & 7.0(2) & 7.0(2) & -- \\
$|B_{MHD}|$~($10^{4}$~Gauss) & 2.1(8) & 0.4(2) & 0.06(2) & 1.(1) & -- \\
$|B_{MCF}|$~($10^{4}$~Gauss) & 80(30) & 14(6) & 5(2) & 30(30) & -- \\
$|B_{\alpha}|$~($10^{4}$~Gauss) & 30-100 & 20-60 & 1.4-2.6 & 70-230 & -- \\
\tableline
\tableline
kT~(keV) &           --          &     --     &    --    &     --            & 1.521(3)  \\
disk~norm. &     --          &     --     &    --    &     --            & $139(1)$ \\
$\Gamma$ &     --         &     --     &    --   &     --           & $2.8(1)$  \\
power-law~norm.  &     --          &     --     &    --   &     --           & $11.0(7)$  \\
$\sigma$~(keV)  &     --          &     --    &    --   &     --            & 0.08(1)  \\
Gauss norm.~$(10^{-3})$ &     --          &     --    &    --   &    --           & 2.2(2)  \\
\tableline
\tableline
\end{tabular}
\vspace*{\baselineskip}~\\ \end{center} 
\tablecomments{Results of fits to the first- and third-order spectra.
  The orders were fit jointly, allowing the third-order spectra to
  have a fiducial continuum, and to set the velocity shifts for Zones
  3 and 4; all other parameters pertain to the first-order spectrum.
  The photoionized wind spectrum is modeled using XSTAR (parameters
  include the column density ${\rm N}_{\rm H}$ and ionizaton paramter
  $\xi$).  Each zone consists of absorption paired with re-emission.
  The re-emission component has zero net shift, and is blurred with
  the ``rdblur'' function to measure dynamical broadening (giving the
  radius $R$, and inclination $\theta$).  The absorption component is
  shifted by $v$ and negative values indicate blue-shifts; the
  re-emission component also carries an overall flux normalization.
  Below the model parameters, estimates of the wind properties are
  given for each zone, including the gas density ($n$), mass outflow
  rate ($\dot{\rm M}_{wind}$), kinetic power ($L_{kin}$), volume
  filling factor ($f$), temperature (${\rm T}_{wind}$), and estimates
  of the magnetic field strength required to launch the wind via MHD
  processes (e.g., MRI) within the disk ($B_{MHD}$) or
  magnetocentrifugal driving ($B_{MCF}$).  The ratio of the mass loss
  rate to mass accretion rate was calculated assuming an efficiency of
  $\eta = 0.1$.  A limit on the field strength predicted by
  $\alpha$-disk models (Shakura \& Sunyaev 1973) is also given
  ($B_{\alpha}$); the range reflects the full range of radii for each
  zone.  Values with an asterisk were frozen.  The overall model gives
  $\chi^{2}/\nu = 1951.2/1572 = 1.241$.  Please refer to
  the text for more details.}
\vspace{-1.0\baselineskip}
\end{footnotesize}
\end{table}
\medskip

\end{document}